\documentclass[aoas,MSNbibl,nameyear,seceqn,dvips]{arximspdf}
\usepackage{dcolumn}
\usepackage{graphicx}

\doi{10.1214/13-AOAS660} 
\volume{7}
\issue{4}
\pubyear{2013}
\firstpage{2229}
\lastpage{2248}

\makeatletter
\newcolumntype{d}[1]{D{.}{.}{#1}}

\newtheorem{theorem}{Theorem}[section]
\newcommand{\convp}{\stackrel{\mathrm{P.}}{\to}}
\newproclaim{definition}{Definition}[section]
\makeatother

\begin{document}
\begin{frontmatter}

\title{An empirical Bayes testing procedure for detecting variants in
analysis of next generation sequencing~data\thanksref{T2}}%
\runtitle{An EB testing procedure for detecting variants}

\begin{aug}
\author[a]{\fnms{Zhigen} \snm{Zhao}\thanksref{m1,t3}\ead[label=e1]{zhaozhg@temple.edu}},
\author[b]{\fnms{Wei} \snm{Wang}\thanksref{m2}\ead[label=e2]{ww42@njit.edu}}
\and
\author[b]{\fnms{Zhi} \snm{Wei}\thanksref{m2}\corref{}\ead[label=e3]{zhiwei@njit.edu}\ead[label=u1,url]{http://ebvariant.sourceforge.net}}
\thankstext{T2}{Supported in part by the National Science Foundation
through major research instrumentation Grant number
CNS-09-58854.}
\thankstext{t3}{Supported by the NSF Grant DMS-12-08735.}
\affiliation{Temple University\thanksmark{m1} and New Jersey Institute
of Technology\thanksmark{m2}}
\runauthor{Z. Zhao, W. Wang and Z. Wei}
\address[a]{Z. Zhao\\
Department of Statistics\\
Temple University\\
346 Speakman Hall\\
1810 N. 13th Street\\
Philadelphia, Pennsylvania 19122\\
USA\\
\printead{e1}}

\address[b]{W. Wang\\
Z. Wei\\
Department of Computer Science\\
New Jersey Institute of Technology\\
GITC 4400, University Heights\\
Newark, New Jersey 07102\\
USA\\
\printead{e2}\\
\phantom{E-mail:\ }\printead*{e3}\\
\printead{u1}}
\end{aug}

\received{\smonth{3} \syear{2012}}
\revised{\smonth{5} \syear{2013}}

%
\begin{abstract}
Because of the decreasing cost and high digital resolution,
next-generation sequencing (NGS) is expected to replace the traditional
hybridization-based microarray technology. For genetics study, the
first-step analysis of NGS data is often to identify genomic variants
among sequenced samples. Several statistical models and tests have been
developed for variant calling in NGS study. The existing approaches,
however, are based on either conventional Bayesian or frequentist
methods, which are unable to address the multiplicity and testing
efficiency issues simultaneously. In this paper, we derive an optimal
empirical Bayes testing procedure to detect variants for NGS study. We
utilize the empirical Bayes technique to exploit the across-site
information among many testing sites in NGS data. We prove that our
testing procedure is valid and \emph{optimal} in the sense of rejecting
the maximum number of nonnulls while the Bayesian false discovery rate
is controlled at a given nominal level.
We show by both simulation studies and real data analysis that our
testing efficiency can be greatly enhanced over the existing
frequentist approaches that fail to pool and utilize information across
the multiple testing sites.
\end{abstract}

%

\end{frontmatter}

\section{Introduction}\label{secintro}
The per-base cost of DNA sequencing has plummeted by  100,000-fold over
the past decade because of the dramatic development in sequencing
technology in the past few years [\citet{Lander2011}]. As a result, this
new or ``next generation'' sequencing (NGS) technology becomes much more
affordable today. With high digital resolution, NGS is expected to
replace the traditional hybridization-based microarray technology [\citet
{Mardis2011}]. For genetics studies, NGS holds the promise to
revolutionize genome-wide association studies (GWAS). In the microarray
era, GWAS mainly addresses common Single Nucleotide Polymorphisms
(SNPs) with minor allele frequency $>$5\%, based upon the common
disease/common variant (CD/CV) hypothesis [\citet{Manolio2009}].
However, the identified common variants explain only a small proportion
of heritability [\citet{Hindorff2009}]. Rare variants therefore have
been hypothesized to account for the missing heritability [\citet
{Bodmer2008,Frazer2009}]. To identify rare variants, a direct and more
powerful approach is to sequence a large number of individuals [\citet
{Li2009}]. This line of thought also implicitly motivates the recent
1000 Genomes Project, which will sequence the genomes of 1200
individuals of various ethnicities by NGS [\citet{Hayden2008}]. It is
expected to extend the catalogue of known human variants down to a
frequency near 1\%. Besides human genetics, NGS is also revolutionizing
genetics in other species. For example, NGS has been used for
genotyping in maize, barley [\citet{Elshire2011}] and rice [\citet
{Huang2009}], accessing allele frequencies genome-wide in Drosophila
[\citet{Turner2011,Zhu2012}], and quantifying strain abundance in yeast
[\citet{Smith2010}]. Because of the small sizes of their genomes,
whole-genome sequencing data for tens or hundreds of samples can be
feasibly generated by one single sequencing run [\citet
{Smith2010,Zhu2012}]. Finally, in cancer genomics, it is interesting to
study the subclonal architecture of tumors. Within a single tumor, that
is, just one individual, there often exists subclones of various sizes
that have distinct somatic mutations. In the case of smaller subclones,
their distinct variants can be present at low frequency when one
sequences the tumor as a whole. To resolve these subclones, one must be
able to accurately identify such low frequency variants and use them to
make inferences about cellular frequency and, thus, subclonal
composition. For such applications, even if one tumor (one sample) is
sequenced as a whole, it actually consists of a pool of heterogeneous
cells from which rare variants are sought.

Thousands of samples need to be sequenced for securing the chance of
finding most rare variants with a frequency $<$1\% [\citet{Li2009}].
A~cost-effective strategy is needed in order to afford very large
sample sizes for finding rare variants.
Similar issues of cost and labor were confronted in the early expensive
stage of GWAS and were circumvented by focusing on small candidate
regions and the use of genomic DNA pooling [\citet
{Sham2002,Norton2004}]. Borrowing the same idea, many targeted
resequencing applications utilizing pooling have been seen in the past
few years [\citet{Nejentsev2009,Out2009,Calvo2010,Momozawa2011}].

Current NGS can generate up to several hundred million reads per run,
which may lead to oversampling with little gain in data quality when
analyzing one sample with a small genome or small targeted genomic
regions. To fully exploit the high-throughput of NGS, nucleotide-based
barcodes have been used to multiplex individual samples [\citet
{Craig2008}]. Different from the aforementioned pooling strategy, this
methodology allows to sequence multiple samples in a single flow cell
while keeping sample identities. However, it should be noted that,
despite the more efficient use of sequencing throughput, multiplexing
techniques still require a large number of individual DNA extractions,
manipulations of reagents, barcoding oligos, PCR reactions and
sequencing library constructions [\citet{Zhu2012}]. For example, in one
of our ongoing projects targeted resequencing 6 Mb genomic regions of
960 human samples, the cost for the library preparation kit (TruSeq
Library Prep${} + {}$NimbleGen Custom EZ Seq Cap Panel) is \$405 per sample
(labor cost not included). We might multiplex 96 samples on one
Illumina HiSeq 2000 lane and get enough sequencing depth per sample
($>$40X/sample). Although the cost for the sequencing step is then
restrained to \$2200 (one lane), the library preparation would cost
dominantly as much as $96*405=\$38\mbox{,}880$, which is not reduced by
multiplexing/barcoding. The library preparation step is cheaper for
whole genome sequencing, as there is no need for capturing targeted
regions. However, the total library preparation cost for multiplexing
tens or more of samples on one lane is still much higher than that for
the sequencing step. In contrast, pooling individuals prior to DNA
extraction and sequencing the pooled DNA without barcodes are very
cost-effective by reducing library preparation cost. As a result, for
population studies where identifying variants and frequencies is the
primary interest rather than knowing which sample the variant came
from, nonindexed multi-sample pools are being widely used to discover
rare variants and/or assess allele frequencies at population level in
Drosophila [\citet{Kolaczkowski2011,Turner2011,Zhu2012}], Anopheles
gambiae [\citet{Cheng2012}], Arabidopsis [\citet{Turner2010}], pig [\citet
{Amaral2011}] and human [\citet{Margraf2011}], among others.

\begin{figure}

\includegraphics{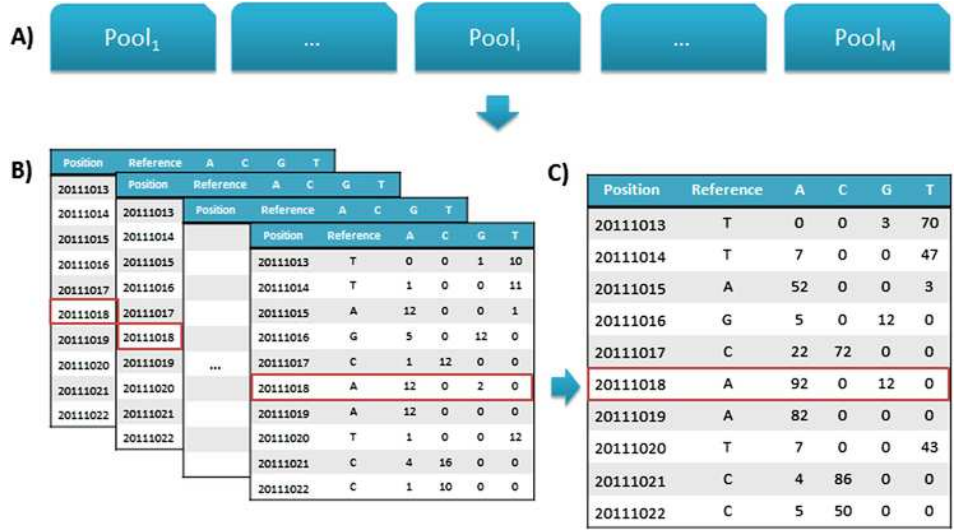}

\caption{Schematic illustration of pooled next generation sequencing
data. \textup{(A)} Suppose M pools are
designed for sequencing and each pool contains N samples. There are two
scenarios for pooled data.
When $M>1$, multiple pool sequencing data are generated. It is possible
that $M>1$ and $N=1$, representing so-called
(individually sequenced) multiple-sample variant call. If $M=1$, it
becomes single pool data. It noted that for
``one'' heterogenous cancer sample, the effective $N$ is larger than~1.
\textup{(B)} For the $i$th pool of $N$ samples, each
nucleotide site is sequenced a random number of times, which yields
different counts of four nucleic acid bases (A, C, G, T)
that make up DNA. \textup{(C)}~An example of pooling $N$ samples at a particular
site. There are two types of alleles:
the reference (major) allele A and the alternative (minor) allele G.
The information is combined from the entire pool of n
individuals.}\label{figintro}
\end{figure}

A schematic example of pooled NGS data is illustrated in Figure~\ref{figintro} assuming there are $M$ pools with $N$ samples in each pool.
For most species, the genetic material DNA is identical at most bases
in a population apart from variations at a small proportion of loci.
Single Nucleotide Variants (SNVs) are the most common DNA sequence
variations occurring when a single nucleotide (A, T, C or G) in the
genome differs between members of a biological species or paired
chromosomes in an individual. SNVs generally exhibit two alleles in a
population. In this particular example, the two alleles, reference
(major) allele and alternative (minor) allele, are A and G,
respectively. Each nucleotide site in each individual chromosome is
sequenced a random number of times. When pooling $N>1$ individuals, the
information of which individual chromosome is represented in a
particular read is lost. In addition, sequencing errors may flip the
original allele into different ones that are observed. It is noted that
when there is only $N=1$ individual in a ``pool'', it represents
so-called (individually sequenced) multiple-sample variant call.
Finally, in the aforementioned cancer genomics studies, because of the
heterogeneity of tumor cell population, the effective $N$ for one
individual tumor sample is believed to be larger than 1.

Identification of genomic variants has become routine after NGS DNA
data are generated. Quite a few tools have been implemented to identify
SNVs. Formally, for a genomic locus, if its minor allele frequency
(MAF) in a population is larger than 0, then we call it a SNV. SNV
detection is a relatively straightforward problem in analysis of
individual data, because the frequency of a candidate allele can be
only 0 (nonvariant), 0.5 (heterozygous) or 1 (alternate homozygous) for
a diploid genome. Several similar conventional Bayesian models have
been used in existing popular tools [\citet
{Li2008}, \citeauthor{Li2009a} (\citeyear{Li2009a,Li2009b}), \citet{McKenna2010}]. The multiplicity issue has been
largely ignored in these conventional Bayesian approaches. Identifying
variants from pooled NGS data is more challenging in that pooled DNA
are sampled from a number of individuals, which consequently will give
rise to variant allele frequencies other than simply 0, 0.5 or 1.
Driven by the need for analysis of increasing amount of pooled NGS
data, quite a few statistical models for the detection of variants from
pooled sequencing data have been developed [\citet
{Druley2009,Bansal2010,Vallania2010,Altmann2011,WeiWangHuLyonHakonarson2011}].
Most existing methods, however, are based on statistical tests from a
frequentist point of view. For example, Wei and colleagues propose a
binomial--binomial model for testing the existence of variants from a
single-pool data [\citet{WeiWangHuLyonHakonarson2011}]. Their
binomial--binomial model provides a unified likelihood function for both
pooled and individual data and has addressed the multiplicity issue.
When there is more than one pool, they employ the partial conjunction
test [\citet{BenjaminiHeller2008}] that at least $u=1$ out of the $M$
hypotheses is false for testing whether a locus is a variant site.
Alternatively, one can also combine individual pool $p$-values by
conducting meta analysis. These frequentist approaches, despite making
few assumptions, fail to pool and utilize information across the
multiple sites that are being tested. Although these approaches are
valid in terms of controlling the FDR at the nominal level, they are
not optimal and powerful in detecting variants of interest. We call an
FDR procedure \emph{valid}, if it controls the FDR at the nominal level
and \emph{optimal}, if it has the smallest false negative rate [FNR,
\citet{GenoveseWasserman2002}] among all valid FDR procedures [\citet
{WeiSunWangHakonsarson2009}]. The optimality issue in multiple
testing has received more and more attention in the past few years
[\citeauthor{SunCai2007} (\citeyear{SunCai2007,SunCai2009}),
\citet{WeiSunWangHakonsarson2009,Wang2010,HeSarkarZhao2011,SunWei2011,XieCaiMarisLi2011}].

Hundreds of thousands or more sites are tested in typical NGS data.
Such high dimensionality imposes great challenges, but can also be a
blessing for inference if handled properly. Empirical Bayesian
approaches, a hybrid of
frequentist and Bayesian methods, become increasingly popular in modern
high-dimensional data inference [\citet{Efron2005talk}].
It enables the frequentists to achieve the Bayesian efficiency in
solving high-dimensional problems [\citet{Efron2010b}].
Assume that the high-dimensional parameters follow some distribution
governed by, for instance, a~few hyperparameters. These hyperparameters
can be estimated reliably via a classical frequentist way. In addition,
empirical Bayesian approaches eliminate the subjective selection of
priors and are generally more robust.

In this article we propose a parametric empirical Bayes testing
procedure for detecting variants in the analysis of high-dimensional
NGS data. When deriving our empirical Bayes procedure, we start from
assuming the hyperparameters are known. Given the known
hyperparameters, we derive a Bayesian decision rule which is optimal in
the sense of detecting the maximum number of variants while the
Bayesian false discovery rate [\citet{SarkarZhouGhosh2008}] is
controlled at a given nominal level. To avoid a subjective choice of
the hyperparameters, we estimate the hyperparameters consistently by
using the method of moments, followed by an empirical Bayes procedure.
Asymptotically, it is guaranteed that the empirical Bayes procedure
mimics the oracle procedure uniformly for all the
hyperparameters.\looseness=-1

In this article we introduce our empirical Bayes testing procedure in
Section~\ref{secmethod}. We present results from simulation studies in
Section~\ref{secsimulation} to demonstrate the superiority of the
proposed procedures in comparison with existing methods. In Section~\ref{secrealdata}, for a case study, we apply the data-driven procedure to
analyze a recent real NGS data set. We present a brief discussion in
Section~\ref{secdiscussion}.
The proof of the theorems are provided in the supplemental article
[\citet{ZhaoWangWeisupp2013}].

The methods developed in this paper have been implemented using Java in
a computationally efficient and user-friendly software package,
EBVariant, as well as an R package, available from \href
{http://ebvariant.sourceforge.net/}{http://ebvariant.sourceforge.net/}.

\section{Statistical models and methods} \label{secmethod}
To discover (rare) variants in a cost-effective way, we consider a
sequencing procedure by pooling a normalized amount of
DNA from multiple samples. Because of a capacity issue, samples may be
distributed and sequenced independently in more than one pools. Without
loss of generality, we assume that there are $M$ pools, each pool with
$N$ individuals (haploids). It is noted that the following proposed
model assumes a general framework and does not require $N>1$. As a
result, when $N=1$ ($N=2$ for a diploid genome), implying each pool has
only one sample, the proposed model is still applicable and will make
an individually sequenced multiple-sample variant call. Suppose that
sequencing covers $p$ sites that are to be tested for variant
candidates. We expect $p$ to be tens or hundreds of thousands for
targeted resequencing, millions for whole-exome sequencing, and
billions for whole-genome sequencing (human). We assume that $K_{ij}$
short reads cover locus $i$ in pool $j$, out of which we observe
$X_{ij}$ reads carry alternative alleles. If there were no sequencing
and mapping errors, we might easily identify variant loci as those with
$X_{ij}>0$. We assume a general sequencing/mapping error $\varepsilon$,
under which the alternative allele will be flipped to one of the other
three alternate alleles, and vice versa. Our goal is to identify single
nucleotide variants (SNVs) that have nonzero minor allele frequencies
in the population.

\subsection{Oracle testing procedure for multiple pools}\label{secmultipool}
We assume that $\theta_{ij}$ is the minor (alternative) allele
frequency (MAF) at the $i$th site in the $j$th pool.
Let $\mu_i\in\{0,1\}$ be the hidden state of whether the $i$th locus is
a SNV. Given $\mu_i=0$, then $\theta_{ij}=0,\forall j=1,2,\ldots, M$.
If $\mu_i=1$, then $\theta_{ij}$'s are nonzero but may vary across
different pools. Following a binomial--binomial model proposed by \citet
{WeiWangHuLyonHakonarson2011}, we assume that the unknown MAF
$\theta_{ij}$ governs $n_{ij}$, the number of haploids in a pool
carrying the alternative alleles, by a binomial model; and that the
unobserved $n_{ij}$ governs its proxy $X_{ij}$ by another binomial
model. Unlike the frequentist approach in \citet
{WeiWangHuLyonHakonarson2011}, we put a prior for $\theta_{ij}$ as
$\psi(\theta_{ij})$ when it is nonzero.\vadjust{\goodbreak} We therefore have a
hierarchical model as follows:
%
%
\begin{equation}
\label{eqnmodelmultipool} \cases{ %
\displaystyle X_{ij}|n_{ij}
\sim b\biggl(K_{ij}, \frac{n_{ij}}{N}(1-\varepsilon )+\biggl(
\frac{N-n_{ij}}{N}\frac{\varepsilon}{3}\biggr)\biggr),
\vspace*{2pt}\cr
n_{ij}|\theta_{ij}\sim b(N,\theta_{ij}),
\vspace*{2pt}\cr
\theta_{ij}|\mu_i\sim(1-\mu_i)
\delta_0+\mu_i\psi(\theta_{ij}),
\vspace*{2pt}\cr
\mu_i\sim \operatorname{Bernoulli}(\pi_0).}
\end{equation}
When there are millions of parameters to be inferred, a common strategy
is to assume that these parameters are drawn from a certain
distribution. We take the parametric approach and assume that $\theta
_{ij}$ follows a uniform distribution $U(0, a)$ with $0<a<1$ when $\mu
_i= 1$.
The corresponding likelihood function of $X_{ij}$ ($i=1,2,\ldots,p,
j=1,2,\ldots,M$) is
%
%
\begin{eqnarray}
\label{eqlikelihood} f(X_{ij}|\theta_{ij},
\mu_i=1)&=&\sum_{n_{ij}=0}^N
\pmatrix{K_{ij} \cr X_{ij}} \biggl(\frac{n_{ij}}{N}(1-
\varepsilon)+\frac
{N-n_{ij}}{N}\frac{\varepsilon}{3}\biggr)^{X_{ij}}
\nonumber\\
&&\hspace*{22pt}{}\times\biggl(1-\biggl(\frac{n_{ij}}{N}(1-\varepsilon)+\frac{N-n_{ij}}{N}
\frac{\varepsilon
}{3}\biggr)\biggr)^{K_{ij}-X_{ij}}\\
&&\hspace*{22pt}{}\times\pmatrix{N \cr {n_{ij}}}\theta_{ij}^{n_{ij}}(1-\theta
_{ij})^{N-n_{ij}}.\nonumber
\end{eqnarray}
When $\mu_i= 0$, the likelihood function becomes
%
%
\begin{equation}
\label{eq10} f(X_{ij}|\mu_i=0)=\pmatrix{K_{ij} \cr {X_{ij}}}
 \biggl(\frac{\varepsilon
}{3}\biggr)^{X_{ij}}\biggl(1-\frac{\varepsilon}{3}
\biggr)^{K_{ij}-X_{ij}}.
\end{equation}

To identify the variants, we test the hypothesis $H_i\dvtx \mu_i=0,
i=1,2,\ldots, p$.
In this multiple-pool scenario, a question remains on how to combine
the data from
multiple pools together.
Wei and colleagues test each single pool separately and combine the
single-pool $p$-values using the Simes' method for testing a partial
conjunction hypothesis [\citet{WeiWangHuLyonHakonarson2011}].
Alternatively, one can conduct the meta-analysis using, for instance,
Fisher's combined probability test [\citet{Fisher1925}]. However, none
of these methods is optimal. We will show in Section~\ref{secsimulation} that these two approaches are conservative in
detecting the variants. The goal of this paper is to construct an
optimal multiple testing procedure by using the Bayesian decision
theory [\citet{HeSarkarZhao2011,SunCai2007}].

Let $\delta_i$ be the 0--1 decision rule corresponding to the $i$th
hypotheses, that is, we reject the hypothesis $H_i$ if $\delta_i=1$.
We consider the loss function
%
%
\begin{equation}
\label{eqloss} L(\bolds{\delta}, \bolds{\mu})=\sum
_i\lambda(1-\mu_i)\delta _i+
\mu_i(1-\delta_i),
\end{equation}
where the tuning parameter $\lambda$ controls the trade-off between the
Type I error and the Type II error.\vadjust{\goodbreak}
Then to minimize the Bayes risk $EL(\bolds{\delta}, \bolds{\mu
})$, we have the Bayesian decision rule $\bolds{\delta}^B=(\delta
_1^B,\ldots,\delta_p^B)$ with
%
%
\begin{equation}
\label{eqdecisionbayes} \delta_i^B=I\biggl(P(
\mu_i=0|\mathbf{X})<\frac{1}{\lambda+1}\biggr).
\end{equation}
Let $fdr_i(\mathbf{X})=P(\mu_i=0|\mathbf{X})$ be the posterior
probability of $\mu_i$ being zero, which is
the local fdr score as given in \citet
{EfronTibshiraniStoreyTusher2001},
\citeauthor{Efron2008} (\citeyear{Efron2008,Efron2010b}).
It can be written as
%
%
\begin{equation}
\label{eq3} \qquad fdr_i(\mathbf{X})=\frac{\pi_0\prod_{j=1}^Mf(X_{ij}|\mu_i=0)}{\pi
_0\prod_{j=1}^Mf(X_{ij}|\mu_{i}=0)+\pi_1\prod_{j=1}^M \int
f(X_{ij}|\theta_{ij})\psi(\theta_{ij})\,d\theta_{ij}}.
\end{equation}
Unlike the two aforementioned approaches, the local fdr score combining
the information across multiple pools proves optimal in the decision
theoretical framework.

The Bayesian decision rule (\ref{eqdecisionbayes}) depends on the
tuning parameter $\lambda$ which, however, is not trivial to set. In
many real applications, of interest is to control certain type I error rates.
False discovery rate (FDR) [\citet{BenjaminiHochberg1995}] is one of
the most popular ones for high-dimensional data. Its recent extensions
include mFDR, which equals $\mathit{FDR}+O(1/p)$ under weak conditions [\citet
{GenoveseWasserman2002}], and positive FDR [\citet{Storey2003}].
Following \citet{SarkarZhouGhosh2008}, we consider the Bayes version
of FDR and FNR (false nondiscovery rate) in the Bayesian framework as follows.

Let $R=\sum_{i=1}^p\delta_i$ and $A=\sum_{i=1}^p(1-\delta_i)$ be the
total number of rejections and acceptances, respectively. Let $V=\sum_{i=1}^p\delta_i(1-\mu_i)$ and
$U=\sum_{i=1}^p\mu_i(1-\delta_i)$ be the number of false rejections and
false acceptances, respectively. Define BFDR and BFNR as
\[
\mathit{BFDR}=E_{\mathbf{X},\bolds{\mu}}\frac{V}{R\vee1},\qquad \mathit{BFNR}=E_{\mathbf{X},\bolds{\mu}}\frac{U}{A\vee1}.
\]

Let $t=\frac{1}{\lambda+1}$ and we rewrite the decision Bayes rule as
$\bolds{\delta}^B(t)=(\delta_1^B(t), \ldots,\break   \delta_p^B(t))$ with
%
%
\begin{equation}
\label{decisionBayes} \delta_i^B(t)=I\bigl(P(
\mu_i=0|\mathbf{X})<t\bigr).
\end{equation}
Then
\[
\mathit{BFDR}\bigl(\bolds{\delta}^B(t)\bigr)=E\frac{\sum_iI(fdr_i(\mathbf
{X})<t)fdr_i(\mathbf{X})}{\sum_iI(fdr_i(\mathbf{X})<t)\vee1},
\]
which is increasing with respect to $t$. As $t\to0$, it converges to
0. When $t\to+\infty$, then
\[
\lim_{t\to+\infty}\mathit{BFDR}\bigl(\bolds{\delta}^B(t)
\bigr)=\frac
{1}{p}E_{m(\mathbf{X})}\sum_ifdr_i(
\mathbf{X})=\pi_0.
\]

Consequently, when $\pi_0>\alpha$, there exists a value $t(\alpha)$
such that the decision Bayes rule controls the BFDR\vadjust{\goodbreak} at $\alpha$ and the
BFDR is greater than $\alpha$ for any $t>t(\alpha)$.
\citet{SunCai2007} and \citet{HeSarkarZhao2011} have shown that this
procedure is optimal in the sense that it yields
the minimal BFNR among all procedures that can control the BFDR at
level $\alpha$.
This optimal rule relies on the cut-off $t(\alpha)$, which
depends on $\alpha$ implicitly.
After deriving the empirical Bayes version of the local fdr scores in
Section~\ref{seceb}, we introduce a data driven procedure to
choose this cutoff in Section~\ref{secdatadriven}.

\subsection{Empirical Bayes estimators}\label{seceb}
The oracle testing procedure defined in Section~\ref{secmultipool}
assumes that the hyperparameters $\pi_0$, $\pi_1$ and $a$ are known.
To avoid a subjective choice of these hyperparameters, we estimate them
using an empirical Bayes approach. To simplify our discussion, we first
explain the estimators for the hyperparameters
for single-pool data. Taking out the pool index $j$, the hierarchical
model for single-pool data becomes
%
%
\begin{equation}
\label{eqnsinglepool} \cases{ %
\displaystyle X_i|n_i
\sim b\biggl(K_i, \frac{n_i}{N}(1-\varepsilon)+\biggl(
\frac
{N-n_i}{N}\frac{\varepsilon}{3}\biggr)\biggr),
\vspace*{2pt}\cr
n_i|\theta_i\sim b(N,\theta_i),
\vspace*{2pt}\cr
\theta_i|\mu_i \sim(1-\mu_i)
\delta_{0}+\mu_iU(0, a),
\vspace*{2pt}\cr
\mu_i\sim \operatorname{Bernoulli}(\pi_0).}
\end{equation}

Define two statistics
%
%
\begin{equation}
\label{eq5} m_1=\frac{\sum_i(X_i/K_i-{\varepsilon}/{3})}{p}
\end{equation}
and
%
%
\begin{eqnarray}
\label{eq6} m_2&=&\frac{1}{p}\sum_i
\bigl(X_i^2 -K_i^2\bigl({\varepsilon^2}/{9}\bigr) - K_i
({\varepsilon}/{3})\bigl(1-({\varepsilon}/{3})\bigr)-K_i\bigl(1-({2\varepsilon
}/{3})\bigr)m_1\nonumber\\
&&\hspace*{224pt}{}-K_i^2({2\varepsilon}/{3})m_1\bigr)\\
&&\hspace*{26pt}{}/{\bigl(\bigl(K_i^2-K_i\bigr)(1-{4\varepsilon}/{3})^2\bigr)}.\nonumber
\end{eqnarray}

\begin{theorem}\label{thmmoment}
Assume the model (\ref{eqnsinglepool}) and the definitions of $m_1$
and $m_2$ in~(\ref{eq5}) and (\ref{eq6}), then
\[
Em_1=\biggl(1-\frac{4\varepsilon}{3}\biggr)\pi_1
\frac{a}{2}
\]
and
\[
Em_2=\frac{N-1}{N}\pi_1\frac{a^2}{3}+
\frac{1}{N}\pi_1\frac{a}{2}.
\]
\end{theorem}
By using the method of moments, we can estimate $a$, $\pi_0$ and $\pi
_1$ as
%
%
\begin{equation}
\label{eq7} \cases{ %
\displaystyle\hat{a}=\frac{3(N(1-{4\varepsilon
}/{3})m_2-m_1)}{2m_1(N-1)},&
\vspace*{2pt}\cr
\displaystyle\hat{\pi}_1=\frac{2m_1}{(1-{4\varepsilon}/{3})\hat{a}},&\quad $\hat{\pi }_0=1-\hat{
\pi}_1.$}
\end{equation}

\begin{theorem}\label{thmconsistent}
Assume that the empirical Bayes estimators of $a$, $\pi_0$ and $\pi_1$
are given by (\ref{eq7}), then
$\hat{a}\convp a$, $\hat{\pi}_0\convp\pi_0$ and $\hat{\pi}_1\convp\pi
_1$, for all $0<a<1, 0<\pi_1<1$.\vspace*{-2pt}
\end{theorem}

The estimation of these hyperparameters borrows information across all
loci and is thus consistent
when the number of loci goes to infinity. This can be viewed as the
blessing of the high dimensionality.
It is noted that the estimation may result in negative estimates of $a$
and $\pi_1$ when $p$ is finite.
For NGS data analysis, people may have certain knowledge about these
unknown parameters. For example, genome-wide $\pi_1$ is believed to be
greater than 0.1\%. We then can set $\hat{\pi}_1$ as 0.1\% if it is
less than 0. Similarly, we may estimate $a$ as $0.01$ if $\hat{a}<0$.
Therefore, we have the truncated estimators for the hyperparameters as
%
%
\begin{eqnarray}
\label{eq4} \cases{ %
\hat{a}^T=\hat{a}I(
\hat{a}>0)+0.01I(\hat{a}<0),
\vspace*{2pt}\cr
\hat{\pi}_1^T = \hat{\pi}_1I(\hat{
\pi}_1>0)+0.001I(\hat{\pi}_1<0),\qquad \hat {
\pi}_0^T=1-\hat{\pi}_1^T.}
\end{eqnarray}
These truncated estimators are still consistent for $\pi_0\in(0, 1)$
and $a\in(0, 1)$.

For the multiple-pool scenario as described in model (\ref
{eqnmodelmultipool}), we assume the observations $X_{ij}, i=1,2,\ldots
, p, j=1,2,\ldots,M$,
share the same marginal distribution. Treating $\{X_{ij}\}$ and $\{
K_{ij}\}$ as
$p\times M$-dimensional vectors, we can estimate $\pi_1$ and~$a$ by
(\ref{eq4}) similarly.
Such estimators converge even faster because of the larger sample size.\vspace*{-2pt}

\subsection{An empirical Bayes testing procedure}\label{secdatadriven}
Section~\ref{secmultipool} has developed an optimal oracle testing procedure.
Section~\ref{seceb} has provided the empirical Bayes estimators for
the parameters $\pi_0$ and $a$ in the testing procedure when they are unknown.
In this section we propose an empirical Bayes testing procedure as follows.\vspace*{-2pt}

\begin{definition}[{[An Empirical Bayes Testing Procedure
(emBayes)]}]\label{ebtesting}
\begin{longlist}[1.]
\item[1.] Estimate $\pi_0$ and $a$ according to (\ref{eq4}).
\item[2.] For the $i$th locus, calculate the local fdr $\widehat{fdr_i(X)}$
by plugging the $\hat{\pi}_1$ and $\hat{a}$ into (\ref{eq3}).
\item[3.] Order $\widehat{fdr_i(X)}$ as $\widehat{fdr_{(1)}(X)}\le\widehat
{fdr_{(2)}(X)}\le\cdots\le\widehat{fdr_{(p)}(X)}$.
\item[4.] Find the maximum $J$ such that $\frac{1}{J}\sum_{i=1}^J\widehat
{fdr_{(i)}(X)}\le\alpha$.
\item[5.] Reject hypothesis $H_{(1)},H_{(2)},\ldots, H_{(J)}$ and accept
the rest.\vspace*{-2pt}
\end{longlist}
\end{definition}

\begin{theorem}\label{thmasymp}
Assume the model (\ref{eqnmodelmultipool}) and the hyperparameters
are estimated as described in Section~\ref{seceb}. Let $\widetilde
{\mathit{BFDR}}$ and $\widetilde{\mathit{BFNR}}$ be the Bayes FDR and FNR of the
empirical Bayes procedure. Then
\[
\widetilde{\mathit{BFDR}}=\mathit{BFDR}_{\mathrm{OR}}+o(1),\qquad \widetilde{\mathit{BFNR}}=\mathit{BFNR}_{\mathrm{OR}}+o(1)
\]
for any $\pi_1\in(0,1)$ and $a\in(0,1)$, where $\mathit{BFDR}_{\mathrm{OR}}$ and
$\mathit{BFNR}_{\mathrm{OR}}$ are the Bayes FDR and FNR of the oracle optimal multiple
testing procedure.\vadjust{\goodbreak}
\end{theorem}

The empirical Bayes procedure was first introduced by \citeauthor{Robbins1951}
(\citeyear{Robbins1951,Robbins1955}), and is also known as a nonparametric
empirical Bayes procedure because the prior is completely unspecified.
Recently, \citet{SunCai2007} and \citet{HeSarkarZhao2011} constructed
optimal nonparametric empirical Bayes multiple testing procedures in
the normal mean setting. In our study, the observation follows a
binomial--binomial model. We put a family of priors with a few
hyperparameters for governing the high-dimensional parameters. The
resultant approach is a parametric empirical Bayes procedure, first
proposed by \citeauthor{EfronMorris1971} (\citeyear{EfronMorris1971,EfronMorris1973,EfronMorris1975}).
Asymptotically, the procedure controls the Bayes
FDR uniformly for all hyperparameter settings. This control is less
stringent than that in the frequentist procedure which requires that
the Bayes FDR be controlled for the class of all point priors on $\theta
$ [\citet{Morris1983b}].
Our empirical Bayes procedure is more robust than the conventional
Bayesian approach which takes a subjective choice of the
hyperparameters. For instance, when setting $\pi_1$ as 0.4\%, the
conventional Bayesian procedure may not control the BFDR if the true
$\pi_1$ is less than 0.4\%, and it may lack power if the true $\pi_1$
is greater than 0.4\%.

\section{Simulation}\label{secsimulation}

We first investigate the numerical performance of the proposed
empirical Bayes procedure (emBayes) using simulated data. Simulation
design follows \citet{WeiWangHuLyonHakonarson2011}, with the
settings: $M=5$ pools, $N=20$ subjects in each pool, the proportion of
alternatives $\pi_1$ varying among 1\%, 0.7\%, 0.3\% and 0.1\%, the MAF
$\psi(\theta_{ij})\sim U(0, a)$ with $a$ being 0.01, 0.02, 0.03 or
0.05, the number of loci $p = 1$ million (1M) or 2 millions (2M), the
sequencing error $\varepsilon=0.01$, and the sequencing coverage $K_{ij}$
following a gamma distribution with mean 30 [\citet{PrabhuPeer2009}].

We compare emBayes with its oracle version, where we use the true
values of $a$ and $\pi_0$, and two frequentist approaches, SNVer and
META. Both SNVer and META test each single pool separately using the
binomial--binomial model. SNVer [\citet
{WeiWangHuLyonHakonarson2011}] combines the single-pool $p$-values
using the Simes' method for testing a partial conjunction hypothesis in
order to get multiple-pool $p$-values. META conducts meta-analysis and
obtains multiple-pool $p$-values as
\[
p^{\mathrm{Pool}}=P\Biggl(\chi^2_{2M}> -2\sum
_{j=1}^M\ln{p_j}\Biggr),
\]
where $\chi^2_{2M}$ is the chi-squared random variable with $2M$
degrees of freedom. Both approaches then employ the BH procedure [\citet
{BenjaminiHochberg1995}] to control FDR.

We evaluate these methods by the number of total rejections (ER), the
number of false rejections (EV)
and the FDR, averaged over 100 replications, at the nominal FDR level
0.05. The results are summarized in Table~\ref{simtab3}. Compared
with SNVer, META
is more conservative and dominated, as indicated by its smaller FDR,
fewer total rejections and fewer true rejections. The results for META
are thus not included in the table.

\begin{table}
\caption{The power and FDR comparison of emBayes, SNVer and the oracle
procedure at the nominal FDR level $5\%$. ER: the number of total
rejections; EV: the number of false rejections; FDR: false discovery rate}\label{simtab3}
\begin{tabular*}{\textwidth}{@{\extracolsep{\fill}}lcccd{1.3}cd{1.3}cd{1.4}@{}}
\hline
&&&\multicolumn{2}{c}{\textbf{emBayes}}&\multicolumn{2}{c}{\textbf{Oracle}}&\multicolumn{2}{c@{}}{\textbf{SNVer}}\\[-6pt]
&&&\multicolumn{2}{c}{\hrulefill}&\multicolumn{2}{c}{\hrulefill}&\multicolumn{2}{c@{}}{\hrulefill}\\
\multicolumn{1}{@{}l}{$\bolds{\pi_1}$} & \textit{\textbf{a}}&\textit{\textbf{p}} & \textbf{ER/EV} & \multicolumn{1}{c}{\textbf{FDR}} &\textbf{ER/EV} &\multicolumn{1}{c}{\textbf{FDR}}&\textbf{ER/EV}&
\multicolumn{1}{c@{}}{\textbf{FDR}}\\
\hline
$\pi_1=1\%$ &0.01&1$M$ &
467/20&0.039&541/27&0.05&277/3.3&0.012\\
&&2$M$ & 1058/52&0.049&1088/55&0.05&563/6.7&0.012\\[3pt]
&0.02&1$M$ & 1464/73&0.05&1467/74&0.05&850/11&0.013\\
&&2$M$ & 2931/144&0.049&2943/147&0.05&1702/22&0.013\\[6pt]
$\pi_1=0.7\%$ &0.01&1$M$ &
295/12&0.038&341/17&0.049&178/2.2&0.012\\
&&2$M$ & 632/28&0.042&682/33&0.049&351/4.2&0.012\\[3pt]
& 0.02&1$M$ & 959/48&0.05&962/48&0.05&533/6.6&0.012\\
&&2$M$ & 1917/94&0.049&1931/97&0.05&1063/13&0.012\\[6pt]
$\pi_1=0.4\%$ &0.01&1$M$ &
132/4.3&0.029&160/7.4&0.046&83/0.9&0.010\\
&&2$M$ & 292/12&0.04&325/16&0.049&170/2.1&0.012\\[3pt]
&0.02&1$M$ & 470/22&0.047&487/24&0.049&257/3.1&0.012\\
&&2$M$ & 971/48&0.049&985/49&0.05&520/6.2&0.012\\[6pt]
$\pi_1=0.1\%$ &0.01&1$M$ &
22/1.1&0.041&26/1.4&0.051&13/0.14&0.01\\
&&2$M$ & 44/1.8&0.032&55/2.4&0.044&26/0.18&0.0068\\[3pt]
&0.02&1$M$ & 73/3&0.036&88/4.4&0.05&45/0.6&0.013\\
&&2$M$ & 153/6&0.035&177/8.3&0.047&90/0.91&0.0099\\
\hline
\end{tabular*}
\end{table}

From Table~\ref{simtab3} we can see that the FDR levels of all three
procedures are controlled at 0.05 asymptotically under all settings
while SNVer is conservative. The power of emBayes is greatly improved
over SNVer.
For instance, when $p=1M$,
$\pi_1=0.4\%$ and $a=0.02$, the numbers of correctly rejected
hypotheses for these two approaches are 470 and 257, respectively. The
number of true rejections is almost doubled.
The emBayes has very comparable, if not the same, performance, compared
with the oracle procedure. The discrepancy is more noticeable
when $\pi_1$ and $a$ are smaller. The reason is that the empirical
Bayes estimators of the hyperparameters
converge slowly near the boundary of the parameter space.

\begin{figure}

\includegraphics{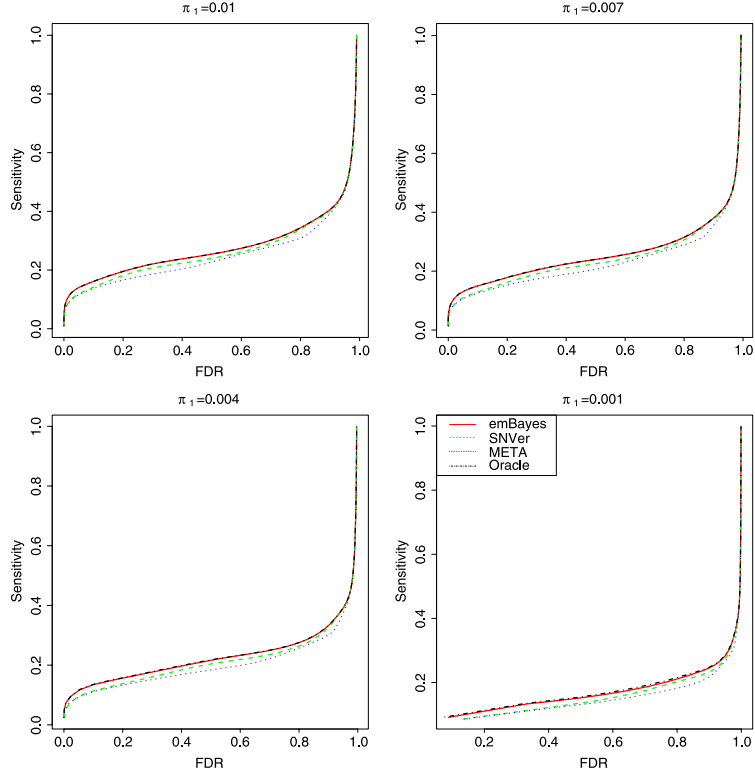}

\caption{ROC curves to compare ranking efficiency of emBayes (red
solid), SNVer (green dashed),
META (blue dotted) and oracle procedure (black-dot dashed) under the
setting of $p=$1$M$ and $a=0.02$
with different proportions of nonnulls.}\label{figroc1}
\end{figure}

In all these simulations, SNVer proves conservative as indicated by
extremely low FDR.
It is tempting to conjecture that the higher power of emBayes is gained
at the price of a higher FDR level. In other words, these two methods
might actually yield similar rankings of the candidate loci and would
demonstrate comparable power at the same empirical FDR level. To
clarify the superiority in terms of prioritizing candidate loci, we
employed ROC curves to illustrate ranking efficiency. Specifically, we
calculated sensitivity as the average proportions of the total number
of true rejections to the total number of nonnulls over the 100
replications. We varied the significance thresholds for identifying up
to 10,000 variants and calculated corresponding FDRs and sensitivities.
The resultant ROC curves of sensitivity versus FDR for emBayes, SNVer,
META and the oracle procedure under the setting of $p={}$1$M$ and $a=0.02$
are shown in Figure~\ref{figroc1}. It is clearly seen that emBayes
dominates SNVer and META. Our proposed empirical Bayes approach can
identify more true variants than the frequentist competitors at the
same FDR levels. For example, when $a=0.02$, $\pi_1=0.1$\% and the FDR
level of 0.1, the numbers of true rejections for emBayes, SNVer, META
and the oracle procedure are 98, 80, 81 and 98, respectively. The
improvement of emBayes over SNVer is as large as $(98-80)/80=22.5$\%.

In summary, our simulation studies show that not only can emBayes
control FDR at nominal level, but, more importantly, it also proves
optimal in terms of power and can detect more variants than its
frequentist alternatives.

\section{Real data analysis}\label{secrealdata}
We also assess the performance of our proposed approach by analyzing a
real NGS data set. In a recent pooled sequencing study, Zhu and
colleagues conducted whole-genome resequencing pools of nonbarcoded
Drosophila melanogaster strains [\citet{Zhu2012}]. The library A
(SRR353364.1) in their study was constructed from a pool of 220 flies
(10 females per strain) and sequenced on a single lane of Illumina
GAIIx platform with 100~bp paired-end reads, leading to an averaged
sequencing depth of 10X. This library was also independently sequenced
by the Drosophila Population Genomics Project (DPGP) (\href
{http://www.dpgp.org/}{http://www.dpgp.org/}). Following the authors,
we utilized this library to evaluate variant call performance.
Specifically, we extracted the genotypes of those 22 strains in the
Library A from the Drosophila Genetic Reference Panel (DGRP)
(\href{http://dgrp.gnets.ncsu.edu}{http://dgrp.gnets.}
\href{http://dgrp.gnets.ncsu.edu}{ncsu.edu}) and used
them as gold standard for estimating False Discovery Rate (FDR).

\begin{figure}

\includegraphics{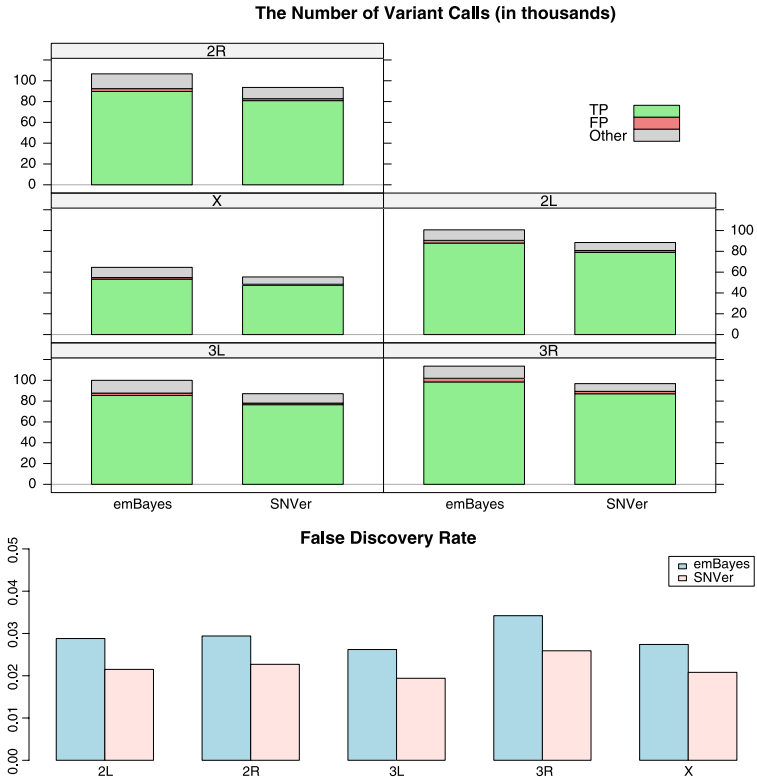}

\caption{Variant call performance. For both methods emBayes and SNVer,
we call variants at the nominal level
$\alpha=0.05$. TP: True Positive; FP: False Positive. TP and FP are the
variants that are called by the method and also have
genotype information available from another data source (DGRP). Other:
the variants called by the method but without genotype
information available from DGRP. FDR: estimated false discovery rate
equal to FP/(TP${}+{}$FP). emBayes calls more variants ($>$10\%)
than SNVer across all five chromosomes. Both methods can control
FDR at the nominal level, while SNVer is more conservative than
emBayes.}\label{figrealdata}
\end{figure}

We downloaded the given bam file, based on which we then called
variants using emBayes and SNVer at the nominal FDR level 0.05. Because
of the large size of Drosophila genome, we analyzed the data separately
for each chromosome. The variant call results are displayed in Figure~\ref{figrealdata}. The emBayes called significantly more varaints than
SNVer across all five chromosomes, with an average of  97,000 variants
per chromosome and the improvement ranging from 13.78\% (Chromosome 2L)
to 17.4\% (Chromosome 3R). Although, as expected, emBayes identified
more variants than SNVer, it is also important to check if these two
methods can control FDR at the prespecified nominal level. The majority
of the called variants (89\%) were found to have their genotype
information available from DGRP, which were then used for estimating
FDR. As we can see from Figure~\ref{figrealdata}, both of the two
methods controlled FDR at the nominal level, while SNVer revealed a
little more conservative than emBayes. Consistent to the simulation
studies, the larger numbers of variants called by emBayes therefore
support its improved power over SNVer.

In summary, the real data analysis confirms that the proposed empirical
Bayesian method, while addressing the multiplicity issue by controlling
FDR, is a more powerful approach by utilizing the global information than
the frequentist approach in detecting variants in NGS study.

\section{Conclusion and discussion}\label{secdiscussion}

This paper has derived an optimal empirical Bayes testing procedure for
detecting variants in analysis of the increasingly popular NGS data.
We utilize the empirical Bayes technique to exploit the across-site
information among the vast amount of testing sites in the NGS data.
We prove that our testing procedure is valid and \emph{optimal} in the
sense of rejecting the maximum number of nonnulls while the marginal
FDR is controlled at a given nominal level.
We show by both simulation studies and real data analysis that our
testing efficiency can be greatly enhanced over the existing
frequentist approaches that fail to pool and utilize information across
the multiple testing sites.

The existing approaches for variant call in NGS study are either
conventional Bayesian models or frequentist tests. Our empirical Bayes
approach can be viewed as a hybrid of the frequentist and Bayesian
methods. It thus enjoys the pros of both and overcomes the cons of
each. Compared to the frequentist approaches, it enjoys the Bayesian
advantage of its capability of pooling information across testing
sites, and therefore is more powerful. In addition, its output local
fdr scores can be used as variant call quality that may be useful in
downstream association analysis [\citet{DayeWeiLi2012}]. Compared to
the conventional Bayesian approaches, it avoids any subjective choice
of prior parameters and estimates them reliably via a classical
frequentist way; it gains multiplicity control by controlling the Bayes
FDR at any designated level uniformly for all the hyperparameters. This
is particularly desirable because tens of thousands or millions of loci
are simultaneously examined in typical NGS experiments. Each user can
choose the false-positive error rate threshold he or she considers
appropriate, instead of just the dichotomous decisions of whether to
``accept or reject the candidates'' provided by most existing methods.

Our current empirical Bayes testing procedure can be extended and
improved in several ways. First, sequencing/mapping error in NGS data
is much more complicated. Due to the heterogeneity of DNA, such as
repeats, duplication and GC content, there could be distinct error
profiles for different genomic regions even if they are sequenced under
the same experimental condition. Instead of assuming a global and
general error rate, we may take and estimate specific and local error
rates empirically from the data for further improving variant call
efficiency. Second, strand bias is an issue observed in many sequencing
platforms but not yet considered in our testing model. We may count and
model ACGT for the forward strand and reverse strand separately, so as
to detect the strand bias and/or allele imbalance issues introduced by
inaccurate mapping or sequencing error. Third, besides single
nucleotide variants (SNVs), there exist small insertions and deletions
(indels). The prevalence and distribution of these indels are quite
different from SNVs. A similar empirical Bayes model but with different
priors may be developed. How to combine them for an overall
multiplicity control while maintaining optimality is not clear. The
recent pooled analysis idea for multiple-testing in GWAS [\citet
{WeiSunWangHakonsarson2009}] may be borrowed and worthy of further
research. We are currently working on these extensions.

\section*{Acknowledgments}
The authors would like to thank the two anonymous referees for their
constructive comments,
which led to a much improved article. The authors thank very much the
area editor
Dr. Karen Kafadar for her valuable time and effort spent on this
submission, without which
the ultimate publication is impossible. Her detailed and specific
comments also helped improve greatly the presentation of the article.

\begin{supplement}[id=suppA]
\stitle{Supplement to ``An empirical Bayes testing procedure for
detecting variants in analysis of next generation sequencing data''\\}
\slink[doi]{10.1214/13-AOAS660SUPP} 
\sdatatype{.pdf}
\sfilename{aoas660\_supp.pdf}
\sdescription{This file contains the technical proof of the theorems.}
\end{supplement}\eject

%

\printaddresses

\end{document}